\documentstyle[emulateapj,apjfonts]{article}
\begin{document}

\submitted{Astrophysical Journal, in press}
\title{Determining the Physical Properties of the B Stars I. Methodology
and First Results}
\author{Edward L. Fitzpatrick}
\affil{Villanova University}
\affil{fitz@ast.vill.edu}
\author{Derck Massa}
\affil{Raytheon ITSS}
\affil{massa@xfiles.gsfc.nasa.gov}

\begin{abstract}

We describe a new approach to fitting the UV-to-optical spectra of B stars 
to model atmospheres and present initial results.  Using a sample of lightly
reddened stars, we demonstrate that the Kurucz model atmospheres can 
produce excellent fits to either combined low dispersion {\it IUE}\/ and 
optical photometry or {\it HST}\/ FOS spectrophotometry, as long as the 
following conditions are fulfilled: 
\begin{enumerate}
\item an extended grid of Kurucz models is employed, 

\item the {\it IUE}\/ NEWSIPS data are placed on the FOS absolute 
flux system using the Massa \& Fitzpatrick (1999) transformation, and 

\item  all of the model parameters  {\em and} the effects of interstellar 
extinction are solved for {\em simultaneously}.   
\end{enumerate}

When these steps are taken, the temperatures, gravities, abundances and 
microturbulence velocities of lightly reddened B0-A0 V stars are determined 
to high precision.  We also demonstrate that the same procedure can be 
used to fit the energy distributions of stars which are reddened by any UV 
extinction curve which can be expressed by the Fitzpatrick \& Massa (1990) 
parameterization scheme.  

We present an initial set of results and verify our approach through
comparisons with angular diameter measurements and the parameters
derived for an eclipsing B star binary.  We demonstrate that the
metallicity derived from the ATLAS~9 fits to main sequence B stars is
essentially the Fe abundance.   We find that a near zero
microturbulence velocity provides the best-fit to all but the hottest
or most luminous stars (where it may become a surrogate for atmospheric
expansion), and that the use of white dwarfs to calibrate UV
spectrophotometry is valid.
  
\end{abstract}


\section{Introduction}\label{intro}

The goal of modeling the observed spectral energy distribution (SED) of
a star is straightforward, namely, to obtain information on the physical
properties of a star from the model atmosphere which fits the
observations best.   After the best-fitting model is found, it is
assumed that the parameters which define the model (effective
temperature, surface gravity, elemental abundances, etc.)  faithfully
represent those of the star itself. The degree to which this key
assumption is valid depends upon both the uniqueness of the fit
(requiring a firm understanding of all observational errors) and the
appropriateness of the physical assumptions used to construct the
models (which often can be verified by ancillary information).   If the
fit provides strong constraints on the model parameters and the model
is considered valid, then this ``continuum fitting'' process provides a
powerful tool for studying stellar properties.

There are two major motivations for modeling the UV/optical SED's of
the main sequence B stars.  First, since 70-to-100\% of the energy of B
stars (with 10000 $< T_{eff} < $30000 K) is emitted in the accessible
spectral region between 1216 \AA\/ and 1 $\mu$m  (see Fig.\ 1 of Bless
\& Percival 1997), even low-resolution UV/optical observations can
place strong constraints on stellar models and should provide robust
determinations of the stellar properties which influence the SED
(particularly $T_{eff}$ and the surface gravity $\log g$).  Second, the
shape of the B star SED's in the UV is also influenced by opacity due
to Fe group elements (Cr, Mn, Fe, Co, Ni), and thus should allow
constraints to be placed on Fe group abundances --- without the
modeling of individual line profiles.  Such results, particularly if
attainable with low-resolution data, would be extremely useful for
studying stellar metallicity patterns when combined with ground-based
determinations of CNO abundances (e.g., Smartt \& Rolleston 1997,
Dufton 1998, Gummersbach et al.\ 1998) since only the coolest of the B
stars have sufficient lines in their optical spectra to allow accurate
Fe group abundances to be derived from the ground.
  
Accessing the information content of the B star UV/optical SED's
requires a set of model atmospheres which reproduce the physical
structure and emergent flux of the B star atmospheres.  The main
sequence B stars are the hottest, most massive, and most luminous stars
whose atmospheric structures are believed to be well-represented by the
simplifying assumptions of local thermodynamic equilibrium (LTE), plane
parallel geometry, and hydrostatic equilibrium.  Thus, although it is
clear that a full non-LTE analysis may be needed for detailed modeling
of specific line profiles, LTE model atmosphere calculations are
expected to reproduce the atmospheric structures and gross emergent
flux distributions of these stars (e.g., Anderson 1985, 1989; Anderson
\& Grigsby 1991; Grigsby, Morrison, \& Anderson 1992; Gummersbach et
al.\ 1998).  As a result, the line-blanketed model atmospheres computed
by R.L. Kurucz, which incorporate the simplifying assumptions listed
above, have become a standard tool in the study of B stars.  These
ATLAS-series models have been used in a wide variety of investigations,
including the determination of fundamental stellar parameters (e.g.,
Code et al.\ 1976); population synthesis (e.g., Leitherer et
al.\ 1996); stellar abundances within the Galaxy (e.g., Geis \& Lambert
1992; Cuhna \& Lambert 1994; Smartt \& Rolleston 1997; Gummersbach et
al.\ 1998) and among the nearby Galaxies (e.g., Rolleston et al.\ 1996
and references therein); the physical structure of atmospheres of the
pulsating $\beta$ Cephei stars (e.g., Cugier et al.\ 1996); the
derivation of interstellar extinction curves (e.g., Aannestad 1995);
the determinations of the distances to binary Cepheids (e.g.,
B\"{o}hm-Vitense 1985); and many others.

Despite the long history of ``continuum fitting'' the UV/optical SED's
of B stars (see, e.g., Code et al.\ 1976, Underhill et al.\ 1979; Remie
\& Lamers 1982; and Malagnini, Faraggiana, \& Morossi 1983; among many
others), several key issues regarding the applicability of the ATLAS
models to the B stars have not yet been addressed quantitatively.
These include the range of stellar properties over which the physical
assumptions and simplifications are valid (or, at least, not
debilitating) and whether --- within that range --- the models truly
reproduce the observations.  Most previous applications of the
continuum fitting technique have relied on ``by eye'' fitting, have
attempted to deduce only $T_{eff}$, and have utilized crude treatments
of interstellar extinction.   Thus it is not clear whether the often
large discrepancies seen in the UV region, in particular, are
attributable to failures in the models or to shortcomings in the
fitting process.

Because of their ubiquity, great diagnostic potential, and relative
ease of calculation, the ATLAS models are a tremendous resource for
stellar astronomy and a quantitative assessment of their range of
applicability is imperative.  Recent significant advances in the
requisite theoretical and observational material make this an ideal
time to undertake such an investigation.  These advances include the
release of the ATLAS~9 models with their greatly improved opacities
(Kurucz 1991), the re-processing of the entire {\it International
Ultraviolet Explorer (IUE)}\/ low-resolution UV spectrophotometry
archives, the derivation of a precise UV calibration by the {\it HST}\/
Faint Object Spectrograph (FOS) team (Bohlin 1996, and see Bless \&
Percival 1997), the derivation of a rigorous transformation between the
{\it IUE}\/ and FOS calibration systems (Massa \& Fitzpatrick 1999,
hereafter MF99), and an improved understanding of the properties of
UV/optical interstellar extinction (Fitzpatrick \& Massa 1990,
hereafter FM; Fitzpatrick 1999, hereafter F99).

This paper is the first in a series whose goals are to quantify the
degree of agreement between the ATLAS model atmospheres and UV/optical
spectrophotometry of B stars and to apply the models to determine the
physical properties of the stars {\em and} the wavelength dependence of
interstellar extinction.  Here we discuss the methodology to be
employed by our study and present some initial results. In \S
\ref{modeling}, we provide an overview of modeling observed energy
distributions and describe the essential ingredients of the fits: a
complete set of model atmosphere calculations, and a quantitative
characterization of the effects of interstellar extinction.  The
section concludes with a description of the mathematical procedure used
to fit the models to the observations and a brief discussion of the
determination of metallicity using the ATLAS~9 models. In \S \ref{data},
we describe the spectrophotometric observations we will be modeling and
\S \ref{results} provides a sample of our results, illustrating the
quality of the fits, the agreement between the derived parameters and
fundamental measurements, and a comparison with previous, independent
abundance measurements.  The potential for future study is also
discussed.  Finally, \S \ref {summary} summarizes our findings and
outline a few remaining technical issues.
 
\section{Modeling the Observed Stellar Fluxes}\label{modeling}

In this section, we provide an overview of the general problem of fitting 
stellar energy distributions, describe the constituents of the model used 
to fit the observations, and discuss how the fitting is accomplished.  

\subsection{Overview of the problem}\label{overview}

The energy distribution of a star as received at the earth, $f_{\lambda
\oplus}$, depends on the surface flux of the star and on the
attenuating effects of distance and interstellar extinction.  This
observed flux at wavelength $\lambda$ can be expressed as:
\begin{eqnarray}
f_{\lambda \oplus} &=& F(\{\alpha\})_{\lambda} \times 
\left(\frac{R}{d}\right)^2 \times 10^{-0.4 A(\lambda)} \\
&=& F(\{\alpha\})_{\lambda} \times \left(\frac{R}{d}\right)^2 \times 
10^{-0.4 E(B-V) [k(\lambda - V) + R(V)]} \label{basic}
\end{eqnarray}
where $F_{\lambda}$ is the emergent flux at the surface of the star (assumed 
to be single), $\{\alpha\}$ is the set of intrinsic physical properties 
which determine the emergent flux from the star, $R$ is the stellar radius, 
$d$ is the distance to the star and $A(\lambda)$ is the total extinction 
along the line of sight.  The factor $(R/d)^2$ is essentially 
equivalent to the square of the angular radius of the star and is often 
written as ($\theta/2)^2$, where $\theta$ is the stellar angular diameter.  

In eq.\ \ref{basic}, the extinction terms are rewritten as quantities
normalized by $E(B-V)$, i.e., the normalized extinction curve $k(\lambda-V) 
= [A(\lambda)-A(V)]/E(B - V)$, and the ratio of selective to total 
extinction in the $V$ band $R(V) \equiv A(V)/E(B-V)$.

\subsection{The models}

We represent the stellar surface fluxes with R.L. Kurucz's ATLAS~9
line-blanketed, LTE, plane-parallel, hydrostatic model atmospheres.
These models are functions of the 4 parameters $\{\alpha\}$ = $\{
T_{eff}, \log g, {\rm [m/H]}, v_t\}$ -- which are the effective
temperature, the log of the surface gravity, the logarithmic metal
abundance relative to the Sun, and the magnitude of the microturbulent
velocity field, respectively.  The standard ATLAS~9 calculations
provide emergent fluxes at 1221 square-binned wavelength points from
90.9 \AA\/ to 160 $\mu$.  The bin widths are typically 10 \AA\/ in the
UV (300 \AA\/ $< \lambda <$ 3000 \AA) and 20 \AA\/ in the optical (3000
\AA $< \lambda <$ 10000 \AA).

Because the fitting procedure must be given access to the entire region
of parameter space which is physically reasonable, it was necessary
extend the standard Kurucz grid, which contains only $v_t = 2$ km
s$^{-1}$ models.  This was accomplished using the ATLAS~9 programs and
input Opacity Distribution Functions (ODFs) to compute additional model
atmospheres and emergent fluxes.  We expanded the grid to include $v_t
= 0, 4, 8$ km s$^{-1}$, [m/H] $= -1.5$, $-1.0$, $-0.5$, 0.0, +0.5
(i.e., from 1/30 to 3$\times$ solar), 36 values of $T_{eff}$ between
9000 and 50000 K, and typically $\sim 10$ values of $\log g$ for each
$T_{eff}$ extending from the Eddington limit to $\log g = 5.0$.  Our
final grid consists of emergent flux distributions for about 7000
models.

In order to calculate a model at a set of parameters $\{T_{eff}^*,
\log g^*, { \rm[m/H]}^*, v_t^*\}$ which do not exactly coincide with a
grid point, an interpolation scheme in all 4 parameters is required.
We adopt the following approach: First, at the three grid values of
$v_t$ closest to $v_t^*$ and the three grid values closest to
[m/H]$^*$, we produce models with $T_{eff}^*$ and $\log g^*$ through
bilinear interpolation of $\log F_{\lambda}$ against $\log T_{eff}$ and
$\log g$.  This yields 9 models.  Next, we quadratically interpolate
$\log F_{\lambda}$ against [m/H] at each of the $v_t$, producing 3
models of the desired $T_{eff}^*$, $\log g^*$, and [m/H]$^*$.  Finally,
we quadratically interpolate $\log F_{\lambda}$ against $v_t$ to obtain
the final model.  Tests of this procedure against models computed
``from scratch'' at $\{T_{eff}^*, \log g^*, { \rm[m/H]}^*, v_t^*\}$
indicate that the errors introduced are negligible.

In addition to fluxes, we produced synthetic Johnson $UBV$ and
Str\"{o}mgren $uvby$ photometry for each model, using the programs of
Buser \& Kurucz (1978) and Relyea \& Kurucz (1978).  The value of the
synthetic V magnitude $V_{syn}$ --- which is used in the current
fitting procedure --- is determined by the following:
\begin{equation}
V_{syn} = - 21.07 -2.5 \times \log \int {f_\lambda S_{\lambda}(V) d\lambda}
\end{equation}
where $f_\lambda$ is a stellar SED (either observed or theoretical) and
$S_{\lambda}(V)$ is the normalized filter sensitivity function for the
$V$ filter from Azusienis \& Straizys (1969).  The integration is
performed over the $V$ filter's range of response (4750 \AA\/ to 7400
\AA).  The normalization zeropoint, $-21.07$, was determined from
synthetic photometry performed on FOS spectrophotometry of seven
calibration stars (Colina \&  Bohlin 1994; Bohlin, Colina, \& Finley
1995) and on ground-based spectrophotometry of HD~172167 (Vega; Hayes
1985).  For these eight stars, equation 3 yields a standard deviation
of 0.01 mag in the difference $V - V_{syn}$, consistent with expected
observational error.

\subsection{Interstellar extinction}\label{reddening}

Correcting stellar SED's for the effects of interstellar extinction is
problematical because the wavelength dependence of extinction is known
to be highly variable spatially.  While average extinction curves have
been defined, there may be few if any sightlines for which some
globally-defined mean curve is strictly appropriate.   Furthermore, the
error introduced by using an inappropriate curve increases in
proportion to the amount of extinction (see Massa 1987).
Consequently, we adopt one of two approaches for modeling extinction,
depending on the color excess of the object.  For lightly reddened
stars ($E(B-V) < 0.1$) we adopt a universal mean curve and solve only
for $E(B-V)$.  For the more reddened stars, we investigate the
feasibility of deriving {\em both} $E(B-V)$ {\em and} the shape of the
extinction curve from the data themselves.  In both cases we utilize an
analytical representation for the extinction wavelength dependence
based on the work of FM.

For the universal mean curve, we use the results of F99.  These are
based on the general UV curve parameterization of FM and the discovery
by Cardelli et al.\ (1989) that the shape of UV extinction curves
correlate with the extinction parameter $R(V)$, defined in \S
\ref{overview}.  F99 presents a mean curve for the case of $R(V) = 3.1$
which (when applied to a stellar energy distribution and convolved with
the appropriate filter responses) reproduces the mean extinction ratios
found in the optical and IR from the Johnson and Str\"{o}mgren
photometric systems.  The curve also joins smoothly with the UV region
(at $\lambda \le 2700 $\AA) via cubic spline interpolation.  The UV
portion of the curve is expressed as a set of three smooth functions
and specified by 6 parameters, all adjusted to their nominal values for
the case $R(V) = 3.1$.
   
For the more heavily reddened stars, we allow for additional free
parameters (i.e, in addition to $E(B-V)$) so that a ``customized''
extinction curve can be derived by solving for some or all of a
modified version of the FM curve parameters.  FM showed that {\em any}
normalized UV extinction curve can be represented by 6 parameters which
describe the general slope and curvature of the curve and the position,
width, and strength of the 2175 \AA\/ bump.  We modify this approach
slightly, in that the linear terms are combined (FM show that these are
{\em functionally} related) and we use a cubic spline to bridge the gap
between the UV at 2700 \AA\/ and the normalized optical curve near 4100
\AA\/ (see F99).  Consequently, the most general UV/optical extinction
curve contains 5 free parameters.  The feasibility of this approach is
demonstrated in \S \ref{results}.

\subsection{Fitting the spectra -- qualitative considerations}

\begin{figure}
\figurenum{1}
%\plotfiddle{figure1.eps}{7.5in}{0}{80}{80}{-230}{-30}
\plotfiddle{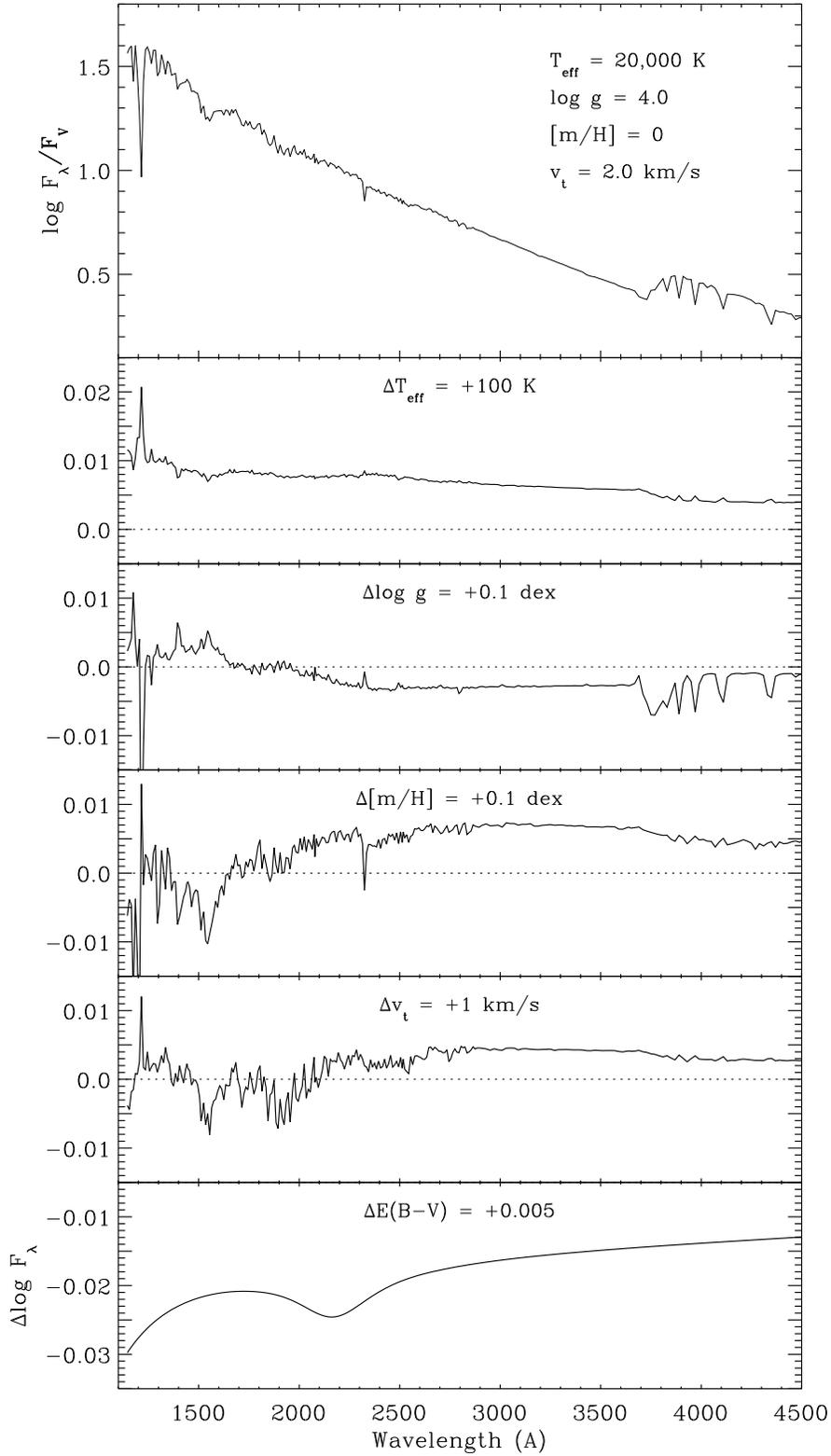}{7.5in}{0}{80}{80}{-230}{0}
\caption{{Representative spectral signatures of the four model
parameters $\{T_{eff}, \log g, [{\rm m/H}], v_t\}$ and $E(B-V)$.  The
top panel shows the normalized flux for a model with $T_{eff} = 20000$
K, $\log g = 4$, [m/H$] = 0$, and $v_t = 2$ km s$^{-1}$.  Each of the
lower panels shows the logarithmic change in the surface flux of the
model when one of the parameters is changed by the amount $\Delta$
indicated in the panel.  The details of the spectral signatures, except
that for $E(B-V)$, are dependent on the basic parameters of the models,
particularly on $T_{eff}$. \label{fig_delta}}}
\end{figure}

The ability to extract the physical parameters of a model accurately
from the observed SED depends on whether changes in those parameters
produce {\em detectable and unique} spectral signatures.  In Figure
\ref{fig_delta}, we illustrate the signatures of the five principal
parameters considered in the fitting process for a typical B star model
flux distribution.  The top panel shows the UV/blue-optical SED for an
ATLAS~9 model with parameters as listed in the figure.  The five lower
panels show the effects on the spectrum caused by changes in the
physical parameters $\{\Delta T_{eff}, \Delta \log g, \Delta {\rm
[m/H]}, \Delta v_t\} = \{+100 {\rm K}, +0.1 {\rm dex}, +0.1, {\rm dex},
+1 {\rm km s}^{-1}\}$ and by a change $\Delta E(B-V) =$ +0.005 mag in
the amount of interstellar extinction present.

Three aspects of the figure are important:

\begin{enumerate}

\item The spectral signatures of the four stellar parameters are
functionally independent, i.e., it is not possible to construct any one
effect from a linear combination of the others.  Consequently, it
should be possible to derive unambiguous and uncorrelated estimates for
each of the parameters from the data.

\item  The signature of interstellar extinction is quite distinct from
those of the model parameters and, in principal, the value of
$E(B-V)$ should be precisely determined by the fitting procedure.

\item The RMS magnitudes of each of the effects in the UV region are of
order 0.01 dex.  Since we will be fitting more than 100 data points in
the UV with typical RMS errors of 3\% per point, uncertainties in the
physical parameters corresponding to the differences shown in the
figure should be well within our grasp.

\end{enumerate}

Note that the strength and shape of each of the spectral signatures
shown in Figure \ref{fig_delta} --- and their degree of independence from
each other --- do vary as a function of the model parameters,
particularly $T_{eff}$.   Therefore, the precision achievable in
estimating the parameters will differ across the wide range of
$T_{eff}$ values appropriate for the B stars.

It might seem surprising that $v_t$ and [m/H] can each be determined
independently of the other, since both directly affect the absorption
lines.  This can be understood in principle, however, from considering
the curve-of-growth, which describes the strength of an absorption
feature as a function of the various relevant physical properties of a
system.  In particular, variations in the number of available absorbers
(e.g., [m/H]) mainly affect weak, unsaturated absorption lines on the
``linear part'' of the curve-of-growth, while variations in line
broadening (e.g., $v_t$) mainly affect the stronger, saturated lines on
the ``flat part'' of the curve-of-growth.  Thus, changes in [m/H] and
$v_t$ have their greatest effects on distinct populations of lines
(although there is a large ``middle ground'' in which both parameters
are important).  This can be seen in Figure \ref{fig_delta} where the
spectral features near 2000 \AA\/ respond dramatically to a change in
$v_t$ and much less so to a change in [m/H].  The inverse can be seen
for features in the 1300--1500 \AA\/ region.
  
The ability to distinguish microturbulence effects from abundance
effects appears to breakdown for cases where $v_t$ is large.  In \S 4
we show several stars for which the combination of a high
microturbulence and a low metallicity reproduce the observed UV opacity
very well --- although their individual values are unlikely to
represent the corresponding stellar properties.  As discussed in \S 4,
this probably indicates that microturbulence is a physically incorrect 
description of the line broadening mechanism in these stars. 
 
 \subsection{Fitting the spectra -- quantitative considerations}

In practice, we fit equation (\ref{basic}) recast into the following form
\begin{eqnarray}
\log f_{\lambda \oplus} &=& \log \left[ \left(\frac{R}{d}\right)^2 \times 
10^{-0.4 A(V)} \right] \nonumber \\  
&& + \log F(T_{eff}, \log g,[\rm m/H],
\it v_t)_{\lambda}\nonumber  \\
&& - 0.4 E(B-V) k(\lambda - V) \label{fit}
\end{eqnarray}
We refer to the leading term as the ``attenuation factor'' because it
is independent of wavelength.  It contains the distance, stellar size
and total extinction information. The second term is the stellar model
which has a non-linear wavelength dependence on the 4 model
parameters.  The final term expresses the magnitude and wavelength
dependence of interstellar extinction and it can represented as either
a simple linear function (in the case of predefined extinction curve)
or a more complex function involving both linear and non-linear terms
(see \S \ref{reddening}).  In general, there can be anywhere between 6
and 11 free parameters, depending on how much flexibility is
incorporated into the description of reddening.

Notice that, since we do not utilize infrared photometry in our
program, we cannot determine the value of $R(V)$ from the fitting
procedure.  Instead, we adopt a typical value (usually $R(V) = 3.1$)
and then incorporate the expected uncertainty into the error estimate
of $\log (R/d)^2$ (an uncertainty in $R(V)$ does not affect the stellar
parameters or $k(\lambda-V)$).
 
The best-fitting model is obtained by solving equation (\ref{fit}) for all 
of the unknown parameters {\em simultaneously}.  We use an iterative, 
weighted non-linear least squares technique (the Marquardt method) to 
adjust the parameters and minimize the $\chi^2$ of the fit.  The parameter
values used to initialize the fitting procedure are determined from the 
spectral type and optical photometry of the star.   The only constraint we
impose on the domain of the model parameters is that $v_t$ and $E(B-V)$ be
$\geq 0$. Note that $E(B-V)$ is simply a scaling factor for $k(\lambda-V)$, 
and we do not utilize $(B-V)$ photometry in the fits at this time.

The degree of accuracy to which any single parameter can be determined
depends on the sensitivity of the observed flux to that parameter and
on the degree of covariance with the other parameters (e.g., see Fig.
\ref{fig_delta}).  Our error analysis incorporates the full
interdependence of the various parameters and includes the effects of
random noise in the data.  All uncertainties quoted in this paper are
1-$\sigma$ internal fitting errors, in the sense that a change in a
particular parameter by $\pm$1-$\sigma$ results in a change in the
total $\chi^2$ by +1 after all the other parameters have been
reoptimized (see Bevington 1969, Chapter 11).

\subsection{Abundances from ATLAS~9 models}\label{atlas_abund}

The definition and impact of the metallicity, [m/H], in the ATLAS~9
models requires special attention.  Different metallicities are
achieved by scaling the abundances of all elements heavier than He by a
single factor relative to the reference solar photosphere abundances
([m/H$]=0$) given by Grevesse \& Anders (1989).  Clearly, a single
parameter is insufficient to describe the results of the complex
physical processes which control the variations in stellar abundance
patterns.  Nevertheless, it is equally clear that such a parameter can
provide a useful diagnostic if it can be determined robustly and with
high precision from relatively low resolution data.

\begin{figure}
\figurenum{2}
\plotfiddle{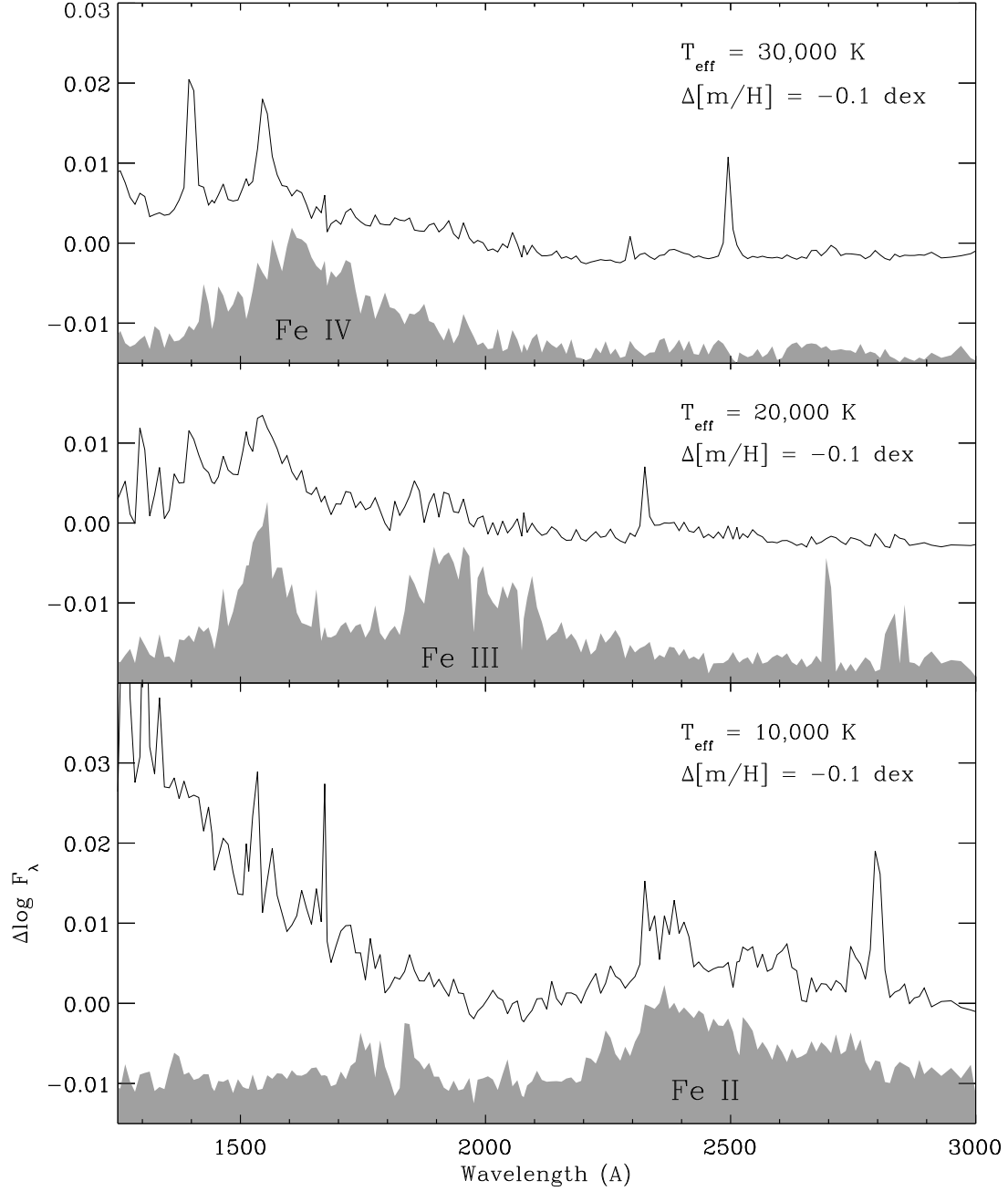}{7.5in}{0}{80}{80}{-240}{-0}
\caption{The spectral signature, $\Delta\log F_{\lambda}$, due to a
change of $\Delta[\rm{m/H}] = -0.1$ dex at three different values of
$T_{eff}$ (solid curve in each panel), compared with a measure of the
strength of the absorption due to various ions of Fe (shaded regions). 
In particular, the shaded regions show the total values of $gf$ for all
the Fe transitions within each ATLAS~9 wavelength bin, arbitrarily scaled
and shifted for display.  The figure demonstrates that, in the UV region
and within the temperature range of the B stars, the overall signature of
changes in the ``metallicity'' of an ATLAS~9 model is mainly due to one
particular element --- Fe.  Other elements contribute within much more
restricted wavelength regions, such as Mg II $\lambda 2800$ (in the
$T_{eff} = 10000$ K case) or Si IV  $\lambda 1400$ and C IV
$\lambda 1550$ (both in the $T_{eff} = 30000$ K  case).  The strong
features in $\Delta \log F_{\lambda}$ at 2325 \AA\/ and 2495 \AA\/ in the
20000 K and 30000 K panels, respectively, are spurious model features
which correspond to no observed stellar lines (see \S 3). \label{fig_iron}}
\end{figure}

The UV spectral region of all main sequence B stars ($10,000 < T_{eff}
< 30,000$ K) is extremely rich in Fe group lines (e.g., Walborn et
al.\ 1995, Roundtree \& Sonneborn 1993), which are obvious even at low
resolution (Swings et al.\ 1973).  The sensitivity of the ATLAS~9
models to changes in [m/H] is dominated by changes in Fe (except over
narrow intervals where the presence of a few strong, isolated silicon
and carbon lines becomes important, see Massa 1989).  This assertion is
justified in Figure \ref{fig_iron} where the solid curves show how of a
$-0.1$ dex change in [m/H] affects the UV SED's for three values of
$T_{eff}$ which span the range of the B stars.  Below each curve, the
shaded regions show the sum of the $gf$ values (arbitrarily scaled and
shifted) for the all spectral lines of Fe IV (top panel), Fe III
(middle panel), and Fe II (bottom panel) within each ATLAS~9
wavelength bin.  Thus, the shaded regions crudely illustrate the
relative strength and general wavelength distribution of the Fe opacity
in the UV.  In the figure, the $\Delta \log F_{\lambda}$ curves have
been paired with Fe ions likely to be dominant, although absorption due
to adjacent ionization stages will also be present.

In each of the panels the opacity signature of the various Fe ions is
strikingly similar to the signature of $\Delta$[m/H].  It is clear that
the value of [m/H] determined by fitting observed SED's with ATLAS~9
models is most heavily influenced by the abundance of Fe.  Therefore we
assume that the value of [m/H] derived from large scale UV absorption
features in B stars is effectively a measurement of [Fe/H].

Several comments regarding the robustness of an [Fe/H] measurement from
fitting the UV SED's of the B stars can be made:

\begin{enumerate}
\item In addition to Fe itself, the other Fe-group elements (Cr, Mn, Fe,
Co, Ni) also contribute to the UV opacity, albeit at a much reduced
level.  However, since all of these elements are thought to have
similar nucleosynthetic origins, their abundances are expected (and
observed) to scale together naturally.   Therefore the ATLAS~9 method
of scaling all abundances by a single parameter is physically
reasonable for the Fe-group elements and does not confuse the
relationship between [m/H] and [Fe/H].

\item The abundances of CNO and the light metals (e.g., Si) certainly
have been observed to vary relative to Fe.  Fortunately, however, the
atmospheric opacity of main sequence B stars is primarily due to
hydrogen, helium and Fe.  Consequently, even if an ATLAS~9 abundance
mix results in too much or too little of CNO or the light metals
relative to the Fe group, the impact on the atmospheric structure and,
hence, the shape of the low resolution energy distribution is both
minimal and localized in wavelength (see the discussion of Vega in \S
4) 
  
\item The low resolution UV spectral signature of Fe results from the
combined opacity of literally thousands of lines.  Therefore, random
uncertainties in individual f-values or line centers tend to cancel
out, as do non-LTE effects in specific lines.
  
\end{enumerate}

Thus, while the study of element-to-element abundance variations
generally requires high resolution stellar data and the computation of
synthetic spectra, Fe abundances in main sequence B stars can be
determined readily from the gross structure of their low resolution UV
spectra.

A second aspect of the ATLAS~9 abundances requires some
clarification.   The models use a reference solar Fe abundance which is
0.12 dex larger than the currently accepted value (see, e.g., Grevesse
\& Noels 1993).  Since, as demonstrated above, Fe dominates the UV
opacity signature from which [m/H] is determined, we adopt the
following transformation:
\begin{equation}
\left[\frac{\rm Fe}{\rm H}\right] \simeq \left[\frac{\rm m}{\rm H}\right] 
+0.12\\
\end{equation}
where [m/H] is the metallicity parameter of the best-fitting ATLAS~9
model and [Fe/H] is the actual logarithmic abundance of Fe relative to
the true solar value.
 
A final comment concerns the $\Delta$[m/H] signature at the cool end of
the B star range.  The lower panel in Figure \ref{fig_iron} (for $T_{eff} =
10000$ K) shows that, while the mid-UV correlation between $\Delta\log
F_{\lambda}$ and the Fe II opacity is unmistakable, the far-UV
sensitivity to $\Delta$[m/H] is larger than might readily be explained
by Fe alone.  Part of this apparent discrepancy is due to changes in
the atmospheric temperature structure which result from changing levels
of opacity.  The far-UV region in these relatively cool models is
extremely sensitive to these changes.  Thus, some of the rapid rise in
$\Delta\log F_{\lambda}$ at short wavelengths is essentially a second
order effect of changing the Fe opacity.  The presence of strong
opacity sources due to elements other than the Fe group also helps
produce the far-UV sensitivity to [m/H].  Numerous lines of C I
are present at wavelengths shortward of $\sim$1500 \AA, along with
strong lines of C II, Si II, and other species.  When this
spectral region is included in the fitting process, the best-fit [m/H]
value will represent a compromise between the abundances of Fe and of
these lighter elements.  For the late B stars, an inability to achieve
good fits between observed and theoretical SED's across the entire UV
spectral region might serve to indicate a more complex abundance
anomaly than can be represented by a single scale factor.  In \S 4 we
will demonstrate this effect in the spectrum of Vega.

\section{The Data}\label{data}

The primary data used in our analysis are NEWSIPS low-dispersion {\it
IUE}\/ UV spectra (Boggess et al.\ 1978a, 1978b; Nichols \& Linsky 1996),
corrected and placed on the {\it HST} - FOS system (Bohlin 1996) using 
the algorithms developed by MF99.  Data from both the
long wavelength (LWR and LWP) and short wavelength (SWP) cameras are
required and the wavelength regions covered are 1978--3000 \AA\/ and
1150--1978 \AA, respectively.   For a few stars, UV-through-optical
spectrophotometry from the FOS is available.  These data typically
extend from 1150 \AA\/ into the optical spectral region.

The UV/optical spectrophotometry (whether from {\it IUE}\/ or FOS) are
resampled to match the wavelength binning of the ATLAS~9 model
atmosphere calculations in the wavelength regions of interest.  Since
the resolution of the {\it IUE}\/ ($\sim$ 4.5 -- 8 \AA) is only
slightly smaller than the size of the ATLAS~9 UV bins (10 \AA), the
characteristics of binned {\it IUE}\/ data do not exactly match those
of ATLAS~9, which may be thought of as resulting from the binning of
very high resolution data.  In essence, adjacent {\it IUE}\/ bins are
not independent of each other, while adjacent ATLAS~9 wavelength points
are independent.  To better match the {\it IUE}\/ and ATLAS~9 data, we
smooth the ATLAS~9 model fluxes in the {\it IUE}\/ wavelength region by
a Gaussian function with a FWHM of 6 \AA.  This slight smoothing
simulates the effects of binning lower resolution data.  No such
smoothing is required when FOS data are used, since they have a much
higher resolution than the {\it IUE}\/ spectra.

For FOS data, the weight assigned to each wavelength bin is derived
from the statistical errors of the data in the bin.  The weights for
the {\it IUE}\/ data are computed likewise, except that an additional,
systematic, uncertainty of 3\% is quadratically summed with the
statistical errors in each bin to account for residual low-frequency
uncertainties not removed by the MF99 algorithms.
   
We give zero weight to the wavelength region 1195 -- 1235 \AA\/ due to
the presence of interstellar Ly $\alpha$ absorption.  Additional points
may be excluded in particular cases based on data quality or the
presense of strong interstellar absorption lines (e.g., Fe II
$\lambda\lambda$2344, 2383, 2600, Mg II $\lambda\lambda$2796, 2802) or
stellar wind features (e.g., Si IV $\lambda$1400, C IV
$\lambda$1550).   Apparently spurious model features are present at
2055, 2325, and 2495 \AA.  These appear as isolated absorption dips in
the models, but correspond to no observed stellar lines.  Their
strengths are temperature dependent, with the 2055 and 2495 \AA\/
features present at model temperatures above $\sim$30000 K and the 2325
\AA\/ feature strongest at model temperatures below 25000 K (this
feature is visible clearly in Figure 1, which also illustrates the
metallicity dependence).  We routinely give zero weight to these three
wavelength bins.

At present, we fit the satellite data and $V$ magnitude data.
Eventually, we will expand this to include Johnson $UBV$ and
Str\"{o}mgren $uvby$ photometry, once the calibration of the synthetic
photometry is more secure (see \S \ref{summary}).  Our sources for the
Johnson and Str\"{o}mgren data are the catalogs of Mermilliod (1987)
and Hauck \& Mermilliod (1990), respectively, when available.

\section{Results and discussion}\label{results}

In this section, we demonstrate our fitting process with several
examples and discuss the consequences of each.  The program stars are
listed in Table \ref{stars_list}.  These stars were selected for three
reasons.  First, as a group, their physical parameters span the full
range of the main sequence B stars.  Second, fundamental measurements
exist for some of them, and these can be used to verify the accuracy of
our results.  Third, some of the stars have been the subject of fine
analyses, so we can compare our derived stellar parameters to previous
estimates, obtained by independent methods.  

\underline{\em BD+33\/$^{\circ}$ 2642}\hspace{0.2cm}  This is a high
latitude post-AGB star and the exciting object of a planetary nebula.
Its spectrum was examined in detail by Napiwotzki, Heber, \& K\"{o}ppen
(1994; hereafter NHK) who found $T_{eff} = 20200$ K, $\log g = 2.9$,
and a peculiar chemical composition where C, N, O, Mg, and Si are
subsolar by about 1 dex and Fe is subsolar by about 2 dex.  This star
is not typical of the type of object at which our research is aimed;
however, it allows us to test the models in two respects.  First, it is
an extreme case of low metallicity.  Second, the FOS data are of
exceptional quality, allowing a critical assessment of how well the
models can fit the details of an actual stellar energy distribution.

\begin{figure}
\figurenum{3}
\plotfiddle{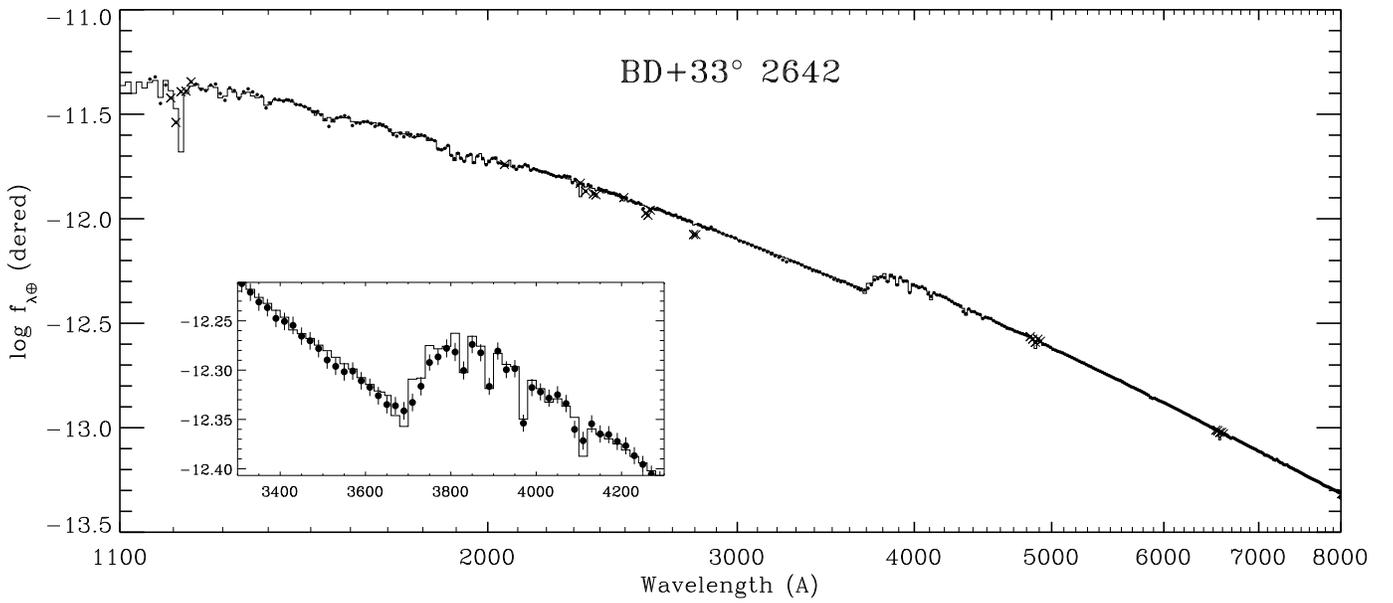}{7.5in}{90}{70}{70}{280}{-0}
\caption{The dereddened {\em FOS} energy distribution of
BD+33$^{\circ}$ 2642 from 1150 \AA\/ to 8000 \AA\/ is shown (solid
circles) along with the best-fitting ATLAS~9 model (solid histogram
line).  Crosses indicate data points excluded from the fit, due to
contamination from interstellar gas-phase absorption lines, spurious
model points, and emission in the stellar H$\alpha$ and H$\beta$
lines.  The inset shows an expanded view of the region surrounding the
Balmer jump.  The parameters of the model are given in Table
\ref{stars_fit}; the $R(V) = 3.1$ extinction curve from F99 was used.
Over the wavelength range shown, the deviations between the model and
the data are typically less than 2$\%$.  \label{fig_bd33}}
\end{figure}

We fit the FOS data for BD+33$^{\circ}$ 2642 using the $R(V) = 3.1$
extinction curve from F99 and we solved for the 4 ATLAS~9 model
parameters, $E(B-V)$, and the attenuation factor.  The result is
presented in Figure \ref{fig_bd33} where we show the reddening-corrected
stellar SED (small filled circles) overplotted with the best-fitting
ATLAS~9 model (solid histogram-style curve).   The fit is so good that
the model and data are nearly indistinguishable over most of the fitted
wavelength range, with typical discrepancies less than 2\% (consistent
with the expected observational errors) except in the regions indicated
by crosses.  These consist of the excluded ATLAS~9 bins discussed in
\S3 above and the regions containing the H$\alpha$ and H$\beta$ lines,
which show evidence for emission in the FOS data.   The model
parameters are listed in columns 2--8 of Table \ref{stars_fit} along with
their 1$\sigma$ internal fitting errors.  The attenuation factor is
represented by the computed angular diameter of the star $\theta$ (in
column 8, in units of milli-arcseconds), as noted in \S2.1.

Our temperature for BD+33$^{\circ}$ 2642 is some 2000 K lower than that
found by NHK.  This is due entirely to the fact that NHK imposed a
value of $E(B-V) = 0.06$, while our self-consistent value is
significantly smaller ($< 0.02$).  (The overcorrection for extinction
in the 2200 \AA\/ region that results from assuming $E(B-V) = 0.06$ is
apparent in Fig. 3 of NHK).  The other parameters derived by NHK will
be adversely affected by the overestimated temperature, but are
nonetheless generally in reasonable agreement with our results.   Our
value of [Fe/H$] = -1.5$ is consistent with the NHK Fe abundance
([Fe/H$] = -2.0\pm0.5$ from {\it IUE}\/ data), but suggests that Fe may
be more similar in abundance to the other elements than emphasized by
NHK.  We find a much lower value of $v_t$ than NHK, who adopted $v_t =
15$ km s$^{-1}$ based on line profile analysis.  Such a large value is
inconsistent with the observed UV opacity and might be an artifact of
their high $T_{eff}$.  It would be interesting to repeat the NHK
analysis using the lower temperature given in Table \ref{stars_fit}.

\underline{\em HV 2274 in the LMC}\hspace{0.2cm}  HV~2274 is a
non-interacting eclipsing binary system in the Large Magellanic Cloud
which consists of two nearly identical early B stars.  The properties
of this system and its usefulness as a distance indicator for the LMC
are discussed by Guinan et al. (1998).  The FOS energy distribution of
HV 2274 (ranging from 1150 to 4800 \AA) was modeled using the program
described here and the results can be seen in Figure 2 of Guinan et
al.  The fitting was performed by solving for the four model
parameters, $E(B-V)$, the attenuation factor, and the six additional
parameters from FM which specify the shape of the UV extinction
curve.   The quality of the fit is comparable to that for
BD+33$^{\circ}$ 2642 discussed above.  The significance of HV~2274 for
our purposes here is that the surface gravities of the stars in the
system are extremely well determined from the properties of the light
and radial velocity curves of the binary.  This affords the opportunity
to test whether the values derived by our approach are reasonable.
This is a critical test since it is not necessarily true that good fits
to energy distributions produce the correct values of the stellar
properties (e.g., see the discussion of HD~38666 and HD~36512 below).
The surface gravity of star A of the binary system is known to be $\log
g = 3.54\pm0.03$.  When the fitting performed with gravity as a free
parameter, a completely consistent value of $\log g = 3.48\pm0.06$ was
determined.  This result gives confidence in the values of the
parameters and suggests that no large systematic effects are present.

\underline{\em HD 38666, 36512, 55857, 31726, 3360, 61831, 90994,}
\underline{\em 209952, 38899, 213398} \hspace{0.2cm}  These form a
sequence of normal Pop I main sequence stars and are representative of
the type of stars on which this study is mainly focused.  This set of
stars demonstrates the quality of fits possible across the entire range
of temperatures typifying the B stars.  In addition, HD~209952 has a
measured angular diameter (Code et al.\ 1976).  Consequently, its
attenuation factor (see eq.\ [\ref{fit}]) can be compared to a direct
measurement.

\begin{figure}
\figurenum{4}
\plotfiddle{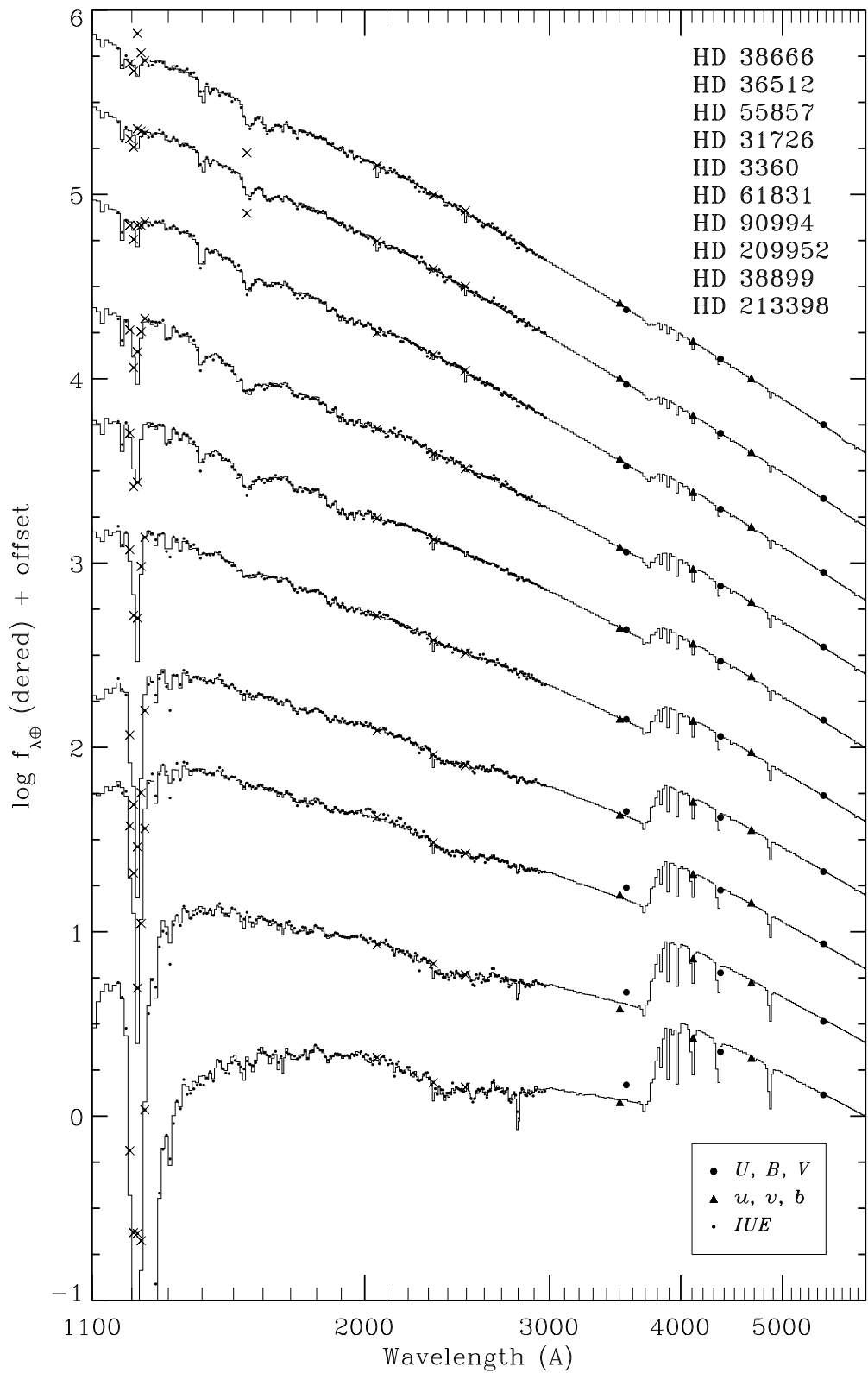}{7.5in}{0}{80}{80}{-260}{-15}
\caption{Sample fits for lightly reddened stars.  Dereddened energy
distributions for 10 main sequence stars ranging from spectral type
O9.5 to A0 V are shown (symbols) along with the best-fitting ATLAS~9
models (solid histogram lines).   Crosses indicate data points excluded
from the fits.  The parameters of the models are given in Table
\ref{stars_fit}; the $R(V)=3.1$ extinction curve from F99 was used.
The Johnson and Str\"{o}mgren photometry data have been converted to
flux units for display only.  The fitting procedure utilizes only the
UV fluxes and the $V$ magnitude.  The $U$ band point appears high in
stars with large Balmer jumps because its broad band straddles the
jump.  The logarithmic fluxes have been arbitrarily displaced
vertically for display purposes.  The offsets applied are, from top to
bottom, 14.22, 13.59, 13.80, 13.41, 12.03, 12.08, 11.78, 10.06, 10.90,
and 10.26  \label{fig_bvs}}
\end{figure}

The observational data for these stars are illustrative of that
available for most of the objects to be studied (at least in the
initial stages of the program) and consist of {\it IUE} UV
spectrophotometry (1150 -- 3000 \AA) and Johnson and Str\"{o}mgren
optical photometry.  The {\it IUE} spectra used here are listed in
Table 3.  The models which best reproduce the observed SED's are shown
in Figure \ref{fig_bvs}, overplotted on the dereddened and arbitrarily
scaled observations.  (The values of the scaling factors are given in
the figure legend.)   We show all the Johnson and Str\"{o}mgren
magnitudes (converted to flux units) for display purposes, but only the
$V$ magnitude was utilized in the fitting procedure.   Data points
excluded from the fit (indicated by crosses) include the region of the
strong Ly $\alpha$ interstellar line, the spurious model points, and
the C IV stellar wind lines in the hottest stars (see \S 3).  As
with BD+33$^{\circ}$ 2642 above, we fit these lightly reddened objects
by assuming the $R(V)= 3.1$ extinction curve from F99 and solving for
the four model parameters, $E(B-V)$, and the attenuation factor.

While the data for these ten stars have lower S/N than for
BD+33$^{\circ}$ 2642, the relative quality of the fits is the same.
I.e., the discrepancies between the observed and model energy
distributions are all consistent with the expected uncertainties in the
data ($\sim$3\% for the {\it IUE}\/ spectra and $\sim$0.015 mag for the
$V$ magnitude).   Note also that the measured value of the angular
diameter $\theta_{observed}$ for HD~209952 (last column of Table
\ref{stars_fit}) agrees with the value derived from the fitting
procedure to within the expected errors.

With the prominent exception of the two hottest stars, HD~38666 and HD
36512, the Fe abundances for the stars in Figure \ref{fig_bvs} are
within about $\pm$0.2 dex of solar.  The mean value for the eight stars
with $T_{eff} < 30000$ K is $\overline{[{\rm Fe/H}]}$ = $-0.07 \pm
0.15$ (s.d.).  The dispersion is larger than expected from the
measurement errors and may indicate a true range in the stellar Fe
abundances.  We will examine this further in the future with a larger
data sample and look for corroborating evidence, such as a systematic
dependence of abundance on spatial location.

The results for HD~38666 and HD~36512 require some comment.
The best-fitting models for these stars not only have very low (and
very implausible) metal abundances, but also very high microturbulence
velocities.   While the combination of these two parameters yields
models which reproduce the observed UV opacity features very well,
nevertheless it seems likely that neither parameter actually
represents a corresponding stellar property.  

We suggest that the large values of $v_t$ signify the presense of a
non-thermal broadening which is {\it not} due to a random velocity
field (i.e., microturbulence) but rather to a velocity gradient in the
photosphere --- indicative of a stellar wind.  The critical difference,
in terms of the effect on the line broadening, is that the continuity
equation (for an atmosphere in whose density decreases with altitude)
requires that a velocity gradient be steepest in the outer layers of
the atmosphere, while microturbulence is depth-independent in the
ATLAS~9 models.  Consequently, the large values of $v_t$ needed to fit
the absorption from strongest lines (i.e., those formed in the
outermost layers of the atmosphere) produce excess absorption by the
less strong lines (formed deeper in the atmosphere).  To compensate for
this effect, the fitting program uses a lower metallicity model to
weaken the latter, resulting in the unexpectedly low [m/H] values
listed in Table \ref{stars_fit}.  If this explanation is correct then
it signals a breakdown in one of the fundamental assumptions of the
ATLAS~9 models, i.e., the applicability of hydrostatic equilibrium, and
casts doubt on the correctness of the other model parameters despite
the good fits to the observed SED's.  We will examine this effect
further with the larger sample, to delineate the region in the
$T_{eff}-\log g$ plane where the values of $v_t$ begin to depart from
the low values characteristic of most of the stars in Figure
\ref{fig_bvs}.

\underline{\em HD~72660 and HD~172167 (Vega)}\hspace{0.2cm}  In recent
years it has been recognized that the ``superficially normal'' early
A stars show a wide range of surface abundance peculiarities.  HD
72660 and Vega show distinctly different types of abundance anomalies
and provide an excellent test of our ability to determine Fe
abundances.  Furthermore, Vega, like HD~209952, has an angular diameter
measurement listed by Code et al.\ (1976).

Vega and HD~72660 have very similar temperatures and surface
gravities.  However, fine spectrum analyses of their optical Fe I
and II lines have shown that Vega has a significantly sub-solar
Fe abundance, [Fe/H$] = -0.55$ (Hill \& Landstreet 1993), while HD
72660 has a super-solar Fe abundance, [Fe/H$] = +0.38$, (Varenne
1999).  Other elements show varying degrees of departure from the solar
standard.  Carbon, for example, is nearly solar in Vega ([C/H$] =
-0.09$) and strongly sub-solar in HD~72660 ([C/H$] = -0.60$).  Only the
coolest of the B stars have adequate optical Fe lines to allow accurate
Fe abundances to be determined from the ground.  Thus, verification of
our approach to determining Fe abundances for Vega and HD72660 will add
considerable confidence to our results for the hotter B stars, where we
must rely exclusively on the UV.

For Vega, we have both an excellent {\it IUE}\/ data set (see Table
\ref{stars_iue}) and high precision optical ground-based
spectrophotometry from Hayes (1985).  The fit to Vega was performed as
for the stars discussed above, by assuming the $R(V)= 3.1$ extinction
curve from F99 and solving for the 4 ATLAS~9 model parameters,
$E(B-V)$, and the attenuation factor.   Both the UV and optical
spectrophotometry (weighted using Hayes' errors) were incorporated in
the fit and the ATLAS~9 models were smoothed by a 50 \AA\/ gaussian in
the optical region to mimic the resolution of the Hayes data.

\begin{figure}
\figurenum{5}
\plotfiddle{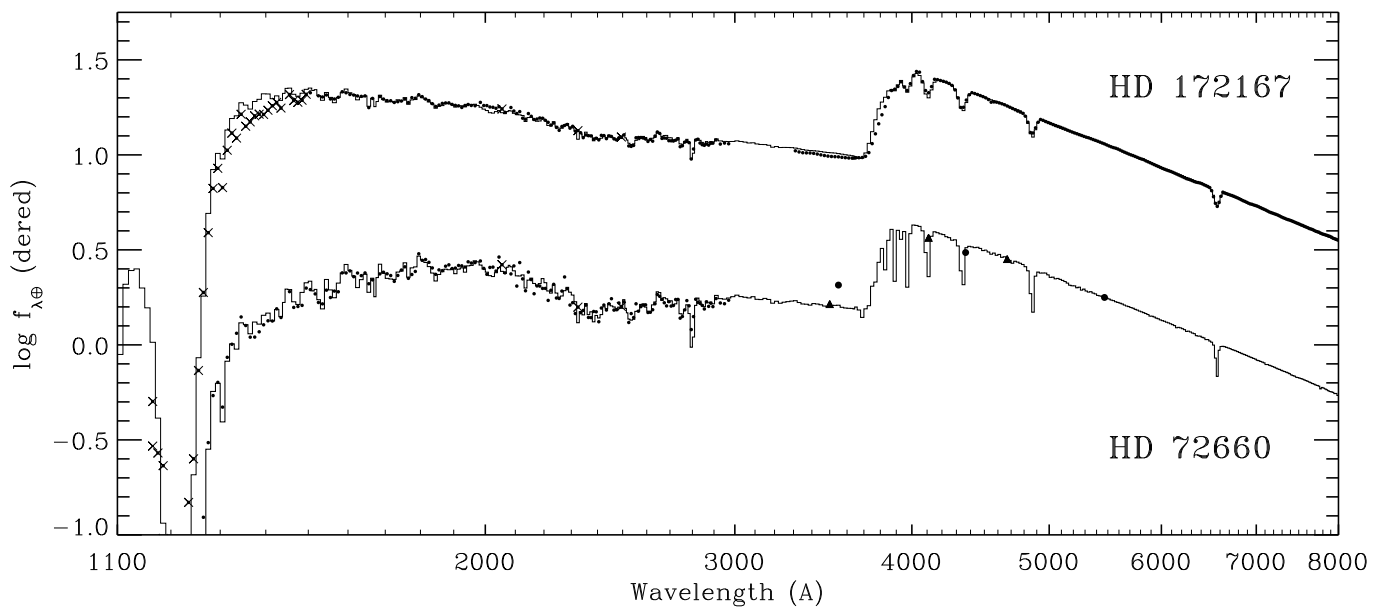}{7.5in}{90}{70}{70}{290}{-0}
\caption{Best-fitting ATLAS~9 models for HD~172167 (Vega; top) and
HD 72660 (bottom).  The symbols are the same as in Fig.\ \ref{fig_bvs}
and the same reddening model was used.  The fit for Vega is relatively
poor between $1200<\lambda<1500$ \AA\/, possibly due to carbon opacity
unaccounted for by the model (see text).  Both stars have similar
temperatures and surface gravities, and their energy distributions
differ primarily because of their different Fe abundances (see Table
\ref{stars_fit}).  \label{fig_avs}}
\end{figure}

It turned out to be impossible to achieve a good fit to Vega
simultaneously over the entire observed spectral region and, in the
end, we gave the region $1200 < \lambda < 1500$ \AA\/ zero weight.  The
results of this fit are shown as the top spectrum in Figure
\ref{fig_avs}, and the derived model parameters --- the most notable of
which is the low value of [m/H] $= -0.73$ --- are listed in Table
\ref{stars_fit}.  As is evident in the figure, the model which reproduces
most of the observed spectrum to within the expected errors but
significantly overestimates the output fluxes between 1200 and 1500
\AA.  The most likely reason for the poor fit in this region is the
complex abundance pattern in Vega.   Because its Fe abundance is so
low, the next largest opacity source in the far UV (C I and II) becomes important.  The model shown in Figure \ref{fig_avs}, which was
optimized to fit the Fe-dominated mid-UV region, underestimates the C
I and II opacity between 1200 and 1500 \AA\/ and thus
overestimates the flux (see \S 2.6).  The value of the Fe abundance for
Vega derived from the fitting procedure, [Fe/H] = --0.61, agrees
extremely well with the result from  Hill \& Landstreet (1993).   A
more detailed discussion of UV/optical energy distribution of Vega can
be found in Castelli \& Kurucz (1994).
  
The fit to HD~72660 was performed in the same manner --- but utilizing
only the $V$ magnitude in the optical region --- and is shown as the
lower spectrum in Figure \ref{fig_avs}.  The fit is good throughout the
entire range, although the overall quality of the observational data is
lower than for Vega so that details in the residuals are not as
apparent.  The parameters of the fit are given in Table \ref{stars_fit};
notable is the large value of [m/H] = +0.25.  Note that we did not need
to disregard the shortest wavelengths in this case, as we did for
Vega.  The opacity is so high in this region due to the elevated Fe
abundance that the concurrent overestimate of the C opacity has minimal
impact.  The derived Fe abundance for HD~72660, [Fe/H]= +0.37, is in
excellent agreement with Varenne's result.
 
\begin{figure}
\figurenum{6}
\plotfiddle{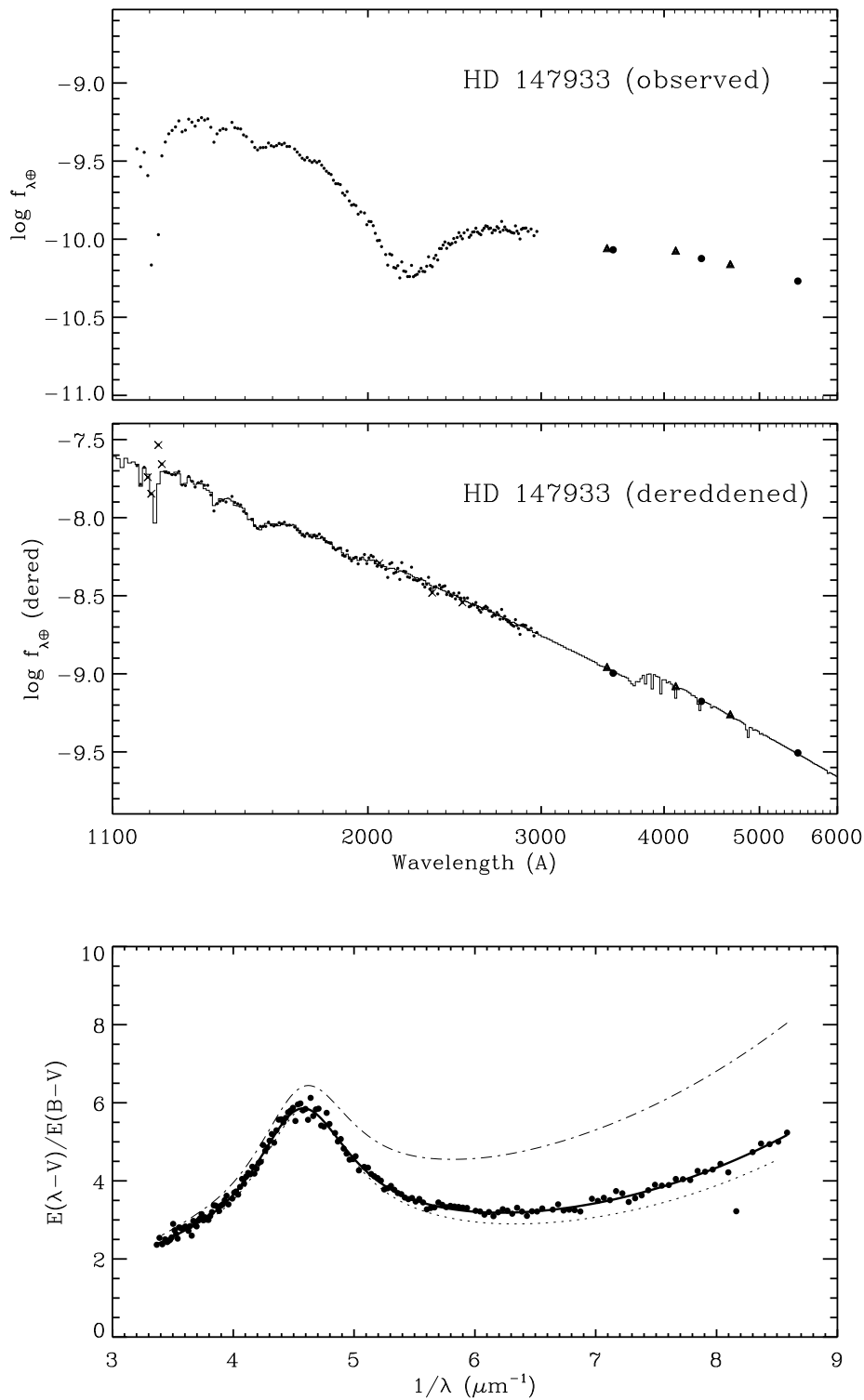}{7.5in}{0}{80}{80}{-240}{-30}
\caption{Best-fitting ATLAS~9 model and UV/optical extinction curve for
HD~147933 ($\rho$ Oph).  The top panel shows the observed {\it IUE}\/
plus optical energy distribution (symbols same as in Fig. 4).  The
middle panel shows the dereddened energy distribution along with the
best-fitting ATLAS~9 model.  The parameters of the model are given in
Table \ref{stars_fit}.  The dereddened energy distribution was also
corrected for the effects of interstellar Ly $\alpha$ absorption,
assuming a column density of $N({\rm H}\/I) = 4.27 \times
10^{21}$ cm$^{-2}$ from Diplas \& Savage 1994.   The lower panel shows
the normalized UV extinction curve determined by the fitting procedure
(thick solid line) overplotted on the actual normalized ratio of the
best-fitting model to the observed (but Ly $\alpha$-corrected) SED
(filled circles).  For comparison, the normalized extinction curve for
HD~147933 derived by FM using the traditional pair method (dotted line)
and the average Galactic curve for $R(V) = 3.1$ from Fitzpatrick 1999
(dash-dot line) are also shown. \label{fig_rho}}
\end{figure}

\underline{\em HD~147933 ($\rho$ Oph)}\hspace{0.2cm}  This early B-main
sequence star is well known for its peculiar UV interstellar extinction
curve.   We use this star to demonstrate the ability of our approach to
extract information on the shape of the interstellar extinction curve
simultaneously with the stellar properties.  The observed SED of HD
147933 is shown in the top panel of Figure \ref{fig_rho}, and consists
of low-resolution {\it IUE}\/ data (see Table \ref{stars_iue}) and
optical photometry.  To fully constrain the models it was necessary to
include not only the {\it IUE} data and the $V$ magnitude in the fit,
but also $B-V$, $b-y$, $m1$, $c1$, and $u-b$.  Inclusion of the colors
required a preliminary calibration of the synthetic photometry, which
was derived from the differences between the observed and calculated
photometry of the other program stars.  We fit the HD~147933 data using
the same six parameters as in the cases above {\em plus} we solve for
the six additional parameters from FM which specify the shape of the UV
extinction curve.  Prior to the fit, the {\it IUE} data outside the
range 1195--1235 \AA\/ (which are always given zero weight) were
corrected for the effects of the strong damping wings of interstellar
Ly $\alpha$ using a column density of $N({\rm H I}) = 4.27\times
10^{21}$ cm$^{-2}$ (Diplas \& Savage 1994).

The best-fitting model and the dereddened energy distribution are shown
in the middle panel of Figure \ref{fig_rho}.  The parameters of the
model, along with $E(B-V)$ and the angular diameter, are given in Table
\ref{stars_fit}.  The bottom panel shows the normalized extinction curve
determined by the fitting program (thick solid line) overplotted on the
actual normalized ratio of the best-fitting model to the observed (but
Ly $\alpha$-corrected) SED (filled circles).  This result is compared
with the ``pair method'' curve for HD~147933 published by FM (dotted
line) and the F99 $R=3.1$ curve (dash-dot line).  The newly derived
curve agrees with the published curve extremely well, lying within the
expected errors for a pair-method curve (Massa \& Fitzpatrick 1986).
This example demonstrates the feasibility of determining extinction
parameters and stellar parameters simultaneously.

\vspace{0.1in}
\section{Summary}\label{summary}

In this section, we summarize our results and expand upon several
noteworthy aspects of the parameters listed in Table \ref{stars_fit}.
 
\begin{enumerate}
\item Our excellent fits to the entire sample of stars listed in Table
\ref{stars_list}, which span the entire range of the main sequence B
stars, are very encouraging.  This is especially true in for the
extremely detailed fit to the FOS spectrum of BD+33$^{\circ}$ 2642.

\item Nevertheless, our results must be verified.  Good fits do not
necessarily mean that the derived physical parameters are correct (see
item 4 below), and verification is essential.  Therefore, it is
gratifying that for those instances where we could compare the
parameters derived from the models to directly measured quantities, the
results agreed.  This includes the angular diameter measurements for
Vega and HD~209952 and the surface gravity measurement for HV2274.

\item Our Fe abundances for Vega and HD~72660 agree with previous
measurements arrived at by completely independent means.  This result
gives us confidence that our metallicity determinations are actually
faithful representations of Fe abundances.  It also justifies our
interpretation of the general metallicity parameter in fits to the
hotter B stars as Fe abundance.  Furthermore the range in our observed
Fe abundances is greater than the errors in the individual
determinations.  This suggests that study of a large sample of lightly
reddened main sequence B stars might reveal patterns in the spatial
distribution of abundances.

\item Our result that $v_t \simeq 0$ km s$^{-1}$ for all but the
hottest main sequence B stars is comforting since it means that our
findings do not have to rely on this poorly defined parameter.  It is
also interesting that $v_t$ appears to assume non-zero values for the
hottest stars in our sample.  Perhaps this is verification of the long
held suspicion that microturbulence is actually a surrogate for
expansion in the early-type stars (e.g., Simonneau 1973; Massa, Shore,
\& Wynne 1992;  Kudritzki 1992).   This suggests that stars which
require large values of $v_t$ may have atmospheres which are not in
hydrostatic equilibrium.  In these cases, the applicability of the
Kurucz ATLAS~9 models is questionable and the parameters derived from
them are suspect.

\item Our results for $\rho$ Oph are very encouraging.  They indicate
that we may be able to extend our analysis in two directions.  One is
the determination of extinction curves for sight lines with far lower
mean reddening than previously possible, since our approach enables us
to characterize the intrinsic flux of a star to far greater precision
than the classic pair method approach, which typically relies on a
coarse grid of standard stars.  The other direction is the
possibility of extending our investigations to include abundance
determinations of more reddened stars, specifically those in the
several open clusters observed by {\it IUE}.

\item Our analysis also provides a powerful verification for the use of
white dwarf models to calibrate UV instruments (Finley \& Koester 1991,
Bohlin 1996). Using a spectrophotometric system calibrated by fitting
white dwarf SEDs to model atmospheres, we obtained excellent fits to B
star SEDs using an independent set of models whose input physics is very
different.  This amounts to a test of the internal consistency of the two
sets of models and their abilities to reproduce observed SEDs.

\end{enumerate}

While the initial results presented here are extremely encouraging, a
few technical issues remain to be resolved.  Leading among these are
the following:
\begin{enumerate}
\item  Verification of the calibration of the synthetic photometry computed
from the ATLAS~9 models.

\item  The impact of binary companions on the parameters derived from
the model fits.

\item  The impact of using the binned ATLAS~9 opacities to
compute the theoretical SED's, instead of using high resolution
synthetic spectrum degraded to the same spectral resolution.

\item  The degree to which the precision of the results falls off with
declining S/N in the observed SED's, as might be relevant for the
application of this technique to extragalactic stars.
\end{enumerate}
We will be examining these issues in future studies as we apply our
results to the several hundred Milky Way stars for which the requisite
data currently exist and begin to examine stars in Local Group
galaxies.
  
In summary, we have introduced and justified a procedure which employs
UV spectrophotometry and optical photometry to {\em simultaneously}
derive the intrinsic stellar parameters $\{T_{eff}, \log g, {\rm [m/H]},
v_t\}$ (where [m/H] is essentially [Fe/H]) of main sequence B stars
and the impact of interstellar extinction on their spectra.  We
demonstrated that our approach produces excellent fits to the
observational data and that reasonable physical parameters are derived
from the models as long as we allow the fitting procedure to access the
full range of parameters and employ the FOS calibration for the UV
data.  This procedure represents a powerful astrophysical tool and
mining the vast {\it IUE}\/ archive of low dispersion B star data will
be extremely rewarding.

\acknowledgements

This program is supported by the Astrophysics Data Program grant NAG5-7113 
to Villanova University and NASA contract NAG5-7372 to Raytheon ITSS.   

%%
%%   BIBLIOGRAPHY
%%

%%
%%   DATA TABLES
%%

\newpage
%\footnotesize
\begin{deluxetable}{lllrrr}
\tablenum{1}
\tablewidth{0pc}
\tablecaption{Program Stars \label{stars_list}}
\tablehead{
\colhead{HD/BD}    & \colhead{Name} & \colhead{Spectral} & \colhead{$V$}   & 
\colhead{$(B-V)$}  & \colhead{$(U-B)$}    \\
\colhead{Number}   & \colhead{}     & \colhead{Type}  &  \colhead{(mag)} &
\colhead{(mag)}         & \colhead{(mag)} }
\startdata
BD+33$^{\circ}$ 2642 &  --              & B2 IVp  & 10.84 & -0.17 & -0.85 \nl
HD \phn\phn3360      &  $\zeta$ Cas     & B2 IV   &  3.66 & -0.20 & -0.85 \nl
HD \phn31726         &  --              & B1 V    &  6.15 & -0.21 & -0.87 \nl
HD \phn36512         &  $\upsilon$ Ori  & B0 V    &  4.62 & -0.26 & -1.07 \nl
HD \phn38666         &  $\mu$ Col       & O9.5 V  &  5.17 & -0.28 & -1.07 \nl
HD \phn38899         &  --              & B9 IV   &  4.90 & -0.07 & -0.17 \nl
HD \phn55857         &  --              & B0.5 V  &  6.11 & -0.24 & -0.99 \nl
HD \phn61831         &  --              & B2.5 V  &  4.84 & -0.20 & -0.65 \nl
HD \phn72660         &  --              & A1 V    &  5.81 &  0.00 &  0.00 \nl
HD \phn90994         &  $\beta$ Sex     & B6 V    &  5.08 & -0.14 & -0.51 \nl
HD 147933            &  $\rho$ Oph      & B2 IV   &  4.59 &  0.23 & -0.57 \nl
HD 172167            &  $\alpha$ Lyr    & A0 Va   &  0.03 &  0.00 & -0.01 \nl
HD 209952            &  $\alpha$ Gru    & B7 IV   &  1.74 & -0.13 & -0.47 \nl
HD 213398            &  $\beta$ PsA     & A0 V    &  4.28 &  0.01 &  0.02 \nl

\enddata
\end{deluxetable}
\normalsize
%\clearpage

%\newpage
\begin{deluxetable}{lrcrcrccc}
\tiny
\tablenum{2} 
\tablewidth{0pc}
\tablecaption{Model Parameters \label{stars_fit}}
\tablehead{
\colhead{Star Name}          & 
\colhead{$T_{eff}$}          & 
\colhead{$\log g$}           & 
\colhead{[m/H]\tablenotemark{1}}         & 
\colhead{[Fe/H]\tablenotemark{1}}             & 
\colhead{$v_t$}              &
\colhead{$E(B-V)$}           &
\colhead{$\theta$\tablenotemark{2}}            &
\colhead{$\theta_{observed}$\tablenotemark{3}}  \\
\colhead{}                   & 
\colhead{(K)}                & 
\colhead{}                   & 
\colhead{}                   & 
\colhead{}                   & 
\colhead{(km s$^{-1}$)}      &
\colhead{(mag)}              &
\colhead{(mas)}              &
\colhead{(mas)}               \\
\colhead{(1)}              & 
\colhead{(2)}              & 
\colhead{(3)}              & 
\colhead{(4)}              & 
\colhead{(5)}              &
\colhead{(6)}              &
\colhead{(7)}              &
\colhead{(8)}              &
\colhead{(9)}               }
\startdata
HD\phn38666     & $31790\pm 500$   & $ 3.85\pm0.08$ & $-0.74\pm0.08$ & $-0.62 $ & $14.7\pm1.5$ & $ 0.021\pm0.004$ & $ 0.10\pm0.01$ & \nodata \nl
HD\phn36512     & $31440\pm 440$   & $ 3.88\pm0.21$ & $-0.82\pm0.09$ & $-0.70 $ & $11.0\pm1.7$ & $ 0.031\pm0.004$ & $ 0.14\pm0.01$ & \nodata \nl
HD\phn55857     & $27740\pm 340$   & $ 3.83\pm0.08$ & $-0.36\pm0.07$ & $-0.24 $ & $ 3.1\pm1.0$ & $ 0.025\pm0.004$ & $ 0.08\pm0.01$ & \nodata \nl
HD147933        & $24070\pm 460$   & $ 4.25\pm0.31$ & $-0.20\pm0.10$ & $-0.08 $ & $ 0.0\pm0.0$ & $ 0.465\pm0.007$ & $ 0.42\pm0.01$ & \nodata \nl
HD\phn31726     & $23330\pm 250$   & $ 4.16\pm0.20$ & $-0.01\pm0.09$ & $+0.11 $ & $ 1.3\pm0.8$ & $ 0.018\pm0.005$ & $ 0.09\pm0.01$ & \nodata \nl
HD\phn\phn3360  & $20900\pm 180$   & $ 3.44\pm0.09$ & $-0.17\pm0.06$ & $-0.15 $ & $ 2.7\pm0.5$ & $ 0.013\pm0.004$ & $ 0.30\pm0.01$ & \nodata \nl
BD+33$^{\circ}$2642 & $18170\pm\phn80$   & $ 2.45\pm0.02$ & $-1.58\pm0.08$ & $-1.46 $ & $ 6.3\pm1.1$ & $ 0.012\pm0.002$ & $ 0.01\pm0.00$ & \nodata \nl
HD\phn61831     & $17900\pm 140$   & $ 3.77\pm0.14$ & $-0.25\pm0.09$ & $-0.13 $ & $ 0.0\pm0.0$ & $ 0.017\pm0.004$ & $ 0.21\pm0.01$ & \nodata \nl
HD\phn90994     & $14420\pm\phn90$ & $ 3.64\pm0.11$ & $-0.16\pm0.05$ & $-0.04 $ & $ 0.0\pm0.0$ & $ 0.006\pm0.004$ & $ 0.22\pm0.01$ & \nodata \nl
HD209952        & $13440\pm\phn50$ & $ 3.76\pm0.11$ & $-0.35\pm0.05$ & $-0.23 $ & $ 0.6\pm0.7$ & $ 0.000\pm0.000$ & $ 1.07\pm0.01$ & $1.02\pm0.07$ \nl
HD\phn38899     & $11040\pm\phn20$ & $ 3.80\pm0.06$ & $-0.20\pm0.03$ & $-0.08 $ & $ 1.1\pm0.5$ & $ 0.000\pm0.000$ & $ 0.30\pm0.01$ & \nodata \nl
HD213398        & $ 9660\pm\phn20$ & $ 4.06\pm0.04$ & $+0.08\pm0.02$ & $+0.20 $ & $ 0.9\pm0.4$ & $ 0.000\pm0.000$ & $ 0.45\pm0.01$ & \nodata \nl
HD172167     & $\phn9600\pm\phn10$ & $ 3.84\pm0.04$ & $-0.73\pm0.06$ & $-0.61 $ & $ 1.9\pm0.4$ & $ 0.000\pm0.000$ & $ 3.26\pm0.01$ & $3.24\pm0.07$ \nl
HD\phn72660  & $\phn9510\pm\phn20$ & $ 3.95\pm0.05$ & $+0.25\pm0.03$ & $+0.37 $ & $ 2.8\pm0.3$ & $ 0.000\pm0.000$ & $ 0.11\pm0.01$ & \nodata \nl
\enddata
\tablenotetext{1}{The quantity [m/H] in column 4 is the ``metallicity'' of the best-fitting ATLAS~9 model.  [Fe/H] in
column 5 represents the
relative abundance of Fe with respect to the currently accepted solar value, computed from [m/H] via equation 5.  See the
discussion in \S 2.6   Note that the results for the stars HD 38666 and HD~36512 are unlikely to represent the true
metallicities of these stars, but rather result from complicating effects due to the onset of strong stellar winds. 
See the discussion in \S 4.}
\tablenotetext{2}{In column 8, the scale factors $(R/d)^2$ determined from the fits have been converted into
angular diameters (in units of milli-arcseconds) as noted in \S 2.1.}
\tablenotetext{3}{The entries in column 9 are directly measured stellar angular diameters from Code et al. 1976.}
\end{deluxetable}
\normalsize
%\clearpage

%\newpage
%\footnotesize
\begin{deluxetable}{llll}
\tablenum{3}
\tablewidth{0pc}
\tablecaption{IUE Spectra Used for Program Stars \label{stars_iue}}
\tablehead{
\colhead{HD}      & \colhead{SWP image}  & \colhead{LWP image} & \colhead{LWR image}  \\
\colhead{Number}  & \colhead{numbers}    & \colhead{numbers}   & \colhead{numbers} }
\startdata
HD \phn\phn3360      &  4316,26364,26365,26366,  & 2931,4866,5041,5042,  & 3812            \nl
                     &  26376,26491,26493,26535, & 6503,6505,6507,6570,  &                 \nl
                     &  26647,26873,26874,26875, & 6692,7528,7529,7530,  &                 \nl
                     &  32869,37984              & 12616,17121           &                 \nl
 HD \phn31726        &  8165               & \nodata          & 7098            \nl
HD \phn36512         &  8164,11155         & \nodata          & 7054,7097,9787  \nl
HD \phn38666         &  14340              & \nodata          & 10954           \nl
HD \phn38899         &  16639              & \nodata          & 12875           \nl
HD \phn55857         &  14339,23117        & 3449             & 10953           \nl
HD \phn61831         &  14309              & \nodata          & 10940           \nl
HD \phn72660         &  44438              & 22854            & \nodata         \nl
HD \phn90994         &  7927,9219,15791    & \nodata          & 6905,7975,12162 \nl
HD 147933            &  6587,6588,6589     & \nodata          & 5639,5640       \nl
HD 172167            &  27024,29864,29867,30548, & 7010,7038,7888,7889,  & \nodata         \nl
                     &  30549,32906,32907        & 10347,10348           &                 \nl
HD 209952            &  26995              & 7009             & \nodata         \nl
HD 213398            &  42704              & 21480            & \nodata         \nl

\enddata
\end{deluxetable}
\normalsize
\clearpage

%%
%%	FIGURES
%%

 \newpage

\end{document}